\documentclass[%
 reprint,
showpacs,preprintnumbers,
 amsmath,amssymb,
 aps,
prb,
]{revtex4-1}

\usepackage{graphicx}% Include figure files
\usepackage{dcolumn}% Align table columns on decimal point
\usepackage{bm}% bold math
\begin{document}

\preprint{APS/Thermalization}

\title{Large K-exciton Dynamics in  GaN Epilayers: the non thermal and thermal regime}% Force line breaks with \\
%\thanks{A footnote to the article title}%

\author{Anna Vinattieri}
 \email{vinattieri@fi.infn.it}
 \affiliation{Dipartimento di Fisica e Astronomia, LENS, CNISM, Universit\`a di Firenze , Italy}%Lines break automatically or can be forced with \\}
\author{Franco Bogani}%
\affiliation{Dipartimento di Energetica, Universit\`a di Firenze, Italy}%
\author{Lucia Cavigli}
 \altaffiliation[Also at ]{Istituto di Fisica Applicata "Nello Carrara", Consiglio Nazionale delle Ricerche, 50019 Sesto Fiorentino (FI), Italy}%Lines break automatically or can be forced with \\
\author{Donatella Manzi}
\author{Massimo Gurioli}
\affiliation{Dipartimento di Fisica e Astronomia, CNISM, LENS, Universit\`a di Firenze , Italy} 
\author{Eric Feltin}
\author{Jean-Fran\c{c}ois Carlin}
\author{Rapha\"el Butt\'e }
\author{Nicolas Grandjean}
\affiliation{Institute of Condensed Matter Physics, Ecole Polytechnique F\'ed\'erale de Lausanne, CH-1015 Lausanne, Switzerland }
\date{\today}% It is always \today, today,
             %  but any date may be explicitly specified

\begin{abstract}
We present a detailed investigation concerning the exciton dynamics in GaN epilayers grown on $c$-plane sapphire substrates, focussing on the exciton formation and the transition from the non-thermal to the thermal regime. The time-resolved kinetics of LO-phonon replicas is used to address the energy relaxation in the excitonic band. From ps time-resolved spectra we bring evidence for a long lasting non-thermal excitonic distribution which accounts for the first 50 ps. Such a behavior is confirmed in different experimental conditions, both when non-resonant and resonant excitation are used. At low excitation power density the exciton formation and their subsequent thermalization is dominated by impurity scattering rather than by acoustic phonon scattering. The estimate of the average energy of the excitons as a function of delay after the excitation pulse provides information on the relaxation time, which describes the evolution of the exciton population to the thermal regime.
%\begin{description}
%\item[Usage]
%Secondary publications and information retrieval purposes.
%\item[PACS numbers]
%May be entered using the \verb+\pacs{#1}+ command.
%\item[Structure]
%You may use the \texttt{description} environment to structure your abstract;
%use the optional argument of the \verb+\item+ command to give the category of each item. 
%\end{description}
\end{abstract}

%\item[PACS numbers]
\pacs{78.30.Fs,78.55.Cr,78.47.jd,71.35.Cc}
% PACS, the Physics and Astronomy
                             % Classification Scheme.
%\keywords{Suggested keywords}%Use showkeys class option if keyword
                              %display desired
\maketitle

%\tableofcontents
\section{\label{Intro}Introduction}
In direct band gap semiconductors the presence of excitonic resonances in optical spectra is considered as a clear marker to assess the sample quality. In the paradigmatic GaAs based systems, excitons are usually probed by means of photoluminescence (PL) experiments and their dynamics is investigated in a restricted  wave vector (k) interval near the $\Gamma$ point (close to k = 0) as a consequence of the radiative recombination process which requires, in the photon emission, both energy and momentum conservation. Therefore,  when an excess energy is used to excite the sample,  the risetime of the PL signal at the exciton energy includes all the information concerning exciton formation and their relaxation towards the k$\approx$0 radiative states, making difficult to extract quantitative information on each process. Even when resonant excitation is used the information on the k$\neq$0 excitons can only be obtained by a careful modeling of the exciton dynamics.\cite{PhysRevB.50.10868}  For semiconductors where excitons strongly interact with the lattice, phonon replicas (PR) of the  k = 0 excitonic band ( the so called zero-phonon line (ZPL)) are present in the PL spectra. Two possible situations can occur depending on the exciton localization/delocalization process in the material. In  disordered/defective materials - where most of the recombinations come from extrinsic contributions - excitons are strongly localized.  The Huang-Rhys (H-R) model  is then commonly used to describe the replica characteristics and to extract the H-R coefficient, which quantifies the exciton interaction with the lattice. This has been extensively done in III-nitrides both in epilayers and quantum confined heterostructures.\cite{leroux99,Kalliakos02,kalliakos:428} In this case the phonon-assisted transitions "replicate" the ZPL shape and contain the same information on the exciton localization dynamics.  Depending on the localization degree a thermal regime can eventually be reached, which requires tens or hundreds of picoseconds. 
\\On the other hand, in samples where exciton localization is weak, the free exciton (FE) thermalization issue is relevant and from the study of the phonon-replica emission we can expect extracting useful information on the exciton distribution over an extended k-range. Moreover, even if the ZPL shows  bound exciton  emission (in non-intentionally doped (nid) GaN epilayers excitons are mainly bound to donor states), the PR bands mostly originate from intrinsic excitonic states,  making negligible the  contribution of the extrinsic band.  The PR bands contain information on the whole (k = 0 and k$\neq$0) exciton population, since the momentum conservation rule is assured by the presence of one or more phonons. In particular longitudinal optical (LO) phonons,  of energy $E_{LO}$, are very effective in determining intense PR-PL bands, which are denominated $n$LO band depending on the number $n$ of LO phonons involved in the emission process. In fact it has been shown that the shape of the 2LO PR of the FE band is proportional to the FE population, while the 1LO PR of the FE band reflects the FE population times the exciton kinetic energy.\cite{Segall68} Therefore phonon replicas have been investigated in different semiconductors to assess both the exciton-phonon interaction and the FE dynamics at large k, which is of interest not only for fundamental physics but also for applications.\cite{Klingshirn} In particular, the study of  the phonon-assisted radiative recombinations provides significant information on the exciton dynamics in II-VI and III-V nitride semiconductors, which show pronounced phonon assisted emission.\cite{GilBook} In III-V nitride structures, most of the reports on PR refer to the steady-state regime.\cite{kovalev96,Buyanova97} We recently showed\cite{cavigli2010} that  a close comparison with existing models\cite{Segall68} for the PR emission in GaN epilayers is relevant provided a free exciton thermalized distribution is achieved and taking into account the complex nature of the exciton band in GaN.  In this way a full reconstruction of the exciton density of states is made possible and the experimental PR energy shift, lineshape and intensity are nicely reproduced by this model.  The k = 0 exciton dynamics has been extensively investigated both at the fs and ps time scale. Typically a 150 ps radiative recombination time and fast dephasing times  (equal to 1-2 ps in high quality samples at low excitation density) are found for the A and B excitons.\cite{monemar2010,hazu02,Ishiguro09}   Nevertheless, despite the fact that in the 1-100 ps range the exciton dynamics is dominated by large-k excitons, the PR recombination kinetics has been scarcely investigated mainly focussing on the quasi-thermal regime.\cite{Hagele99} But even if the exciton-acoustic phonon interaction is quite strong, the thermal regime is reached after a finite time. This implies the presence of a non-thermal regime, which has not been experimentally investigated yet. Only theoretical results concerning the exciton formation and the approach to thermal equilibrium are reported in literature.\cite{Kuhm11,Kokolakis03,Ridley88} In this study we aim at filling this gap and therefore we explore both the non-thermal and the thermal regime. From picosecond time-resolved experiments performed on nid GaN epilayers grown on c-plane sapphire substrates, we show that the achievement of a hot thermal distribution of excitons requires a few tens of ps, depending on the excitation energy, the lattice temperature and the excitation power density. The heating process is also studied by resonantly exciting the k = 0 FE (X$_{\mathrm{A}}$ exciton). As a consequence of the different wavevector intervals explored by the PR and the ZPL emission, we provide insights on the exciton dynamics in an extended region near the  $\Gamma$-point of the first Brillouin zone.

\section{\label{sec:experimental}Experimental}

The investigated samples are a 11 $ \mu $m thick nid wurtzite GaN epilayer grown by hybrid vapor phase epitaxy  (HVPE) on a $c$-plane sapphire substrate\cite{martin06} (sample A) and a 3 $ \mu $m thick nid wurtzite GaN epilayer grown by metal organic vapor phase epitaxy  (MOCVD) on a $c$-plane sapphire substrate (sample B).  The threading dislocation density is  as low as $2\times10^{8}$ cm$^{-2}$ in the HVPE sample and lower than $1\times10^{9}$   cm$^{-2}$  in the MOVPE epilayer with a not intentional $n$-type doping $\approx 10^{17}/$cm$^{-3}$. Measurements were performed with the sample placed in a closed-cycle cryostat varying the temperature from 10 to 50 K for time-resolved experiments. Reflectivity  measurements at normal incidence were carried out using a CW Xe lamp. Time-integrated photoluminescence (TI-PL) measurements were performed under non-resonant excitation by a frequency-doubled ps dye laser at 300 nm, and  the collected light was detected by a cooled silicon charge coupled device camera after dispersion through a 50 cm flat-field spectrometer with a spectral resolution of 1 meV. Time-resolved photoluminescence (TR-PL) measurements were performed using a frequency-doubled mode-locked Ti:sapphire laser pumped by a continuous wave Ar$^{+}$ laser providing 1.2 ps pulses at a repetition rate of 81 MHz. The PL was dispersed through a 30 cm flat-field monochromator and detected by a streak camera apparatus with a 3 ps time resolution.
\begin{figure}	\includegraphics[scale=0.5]{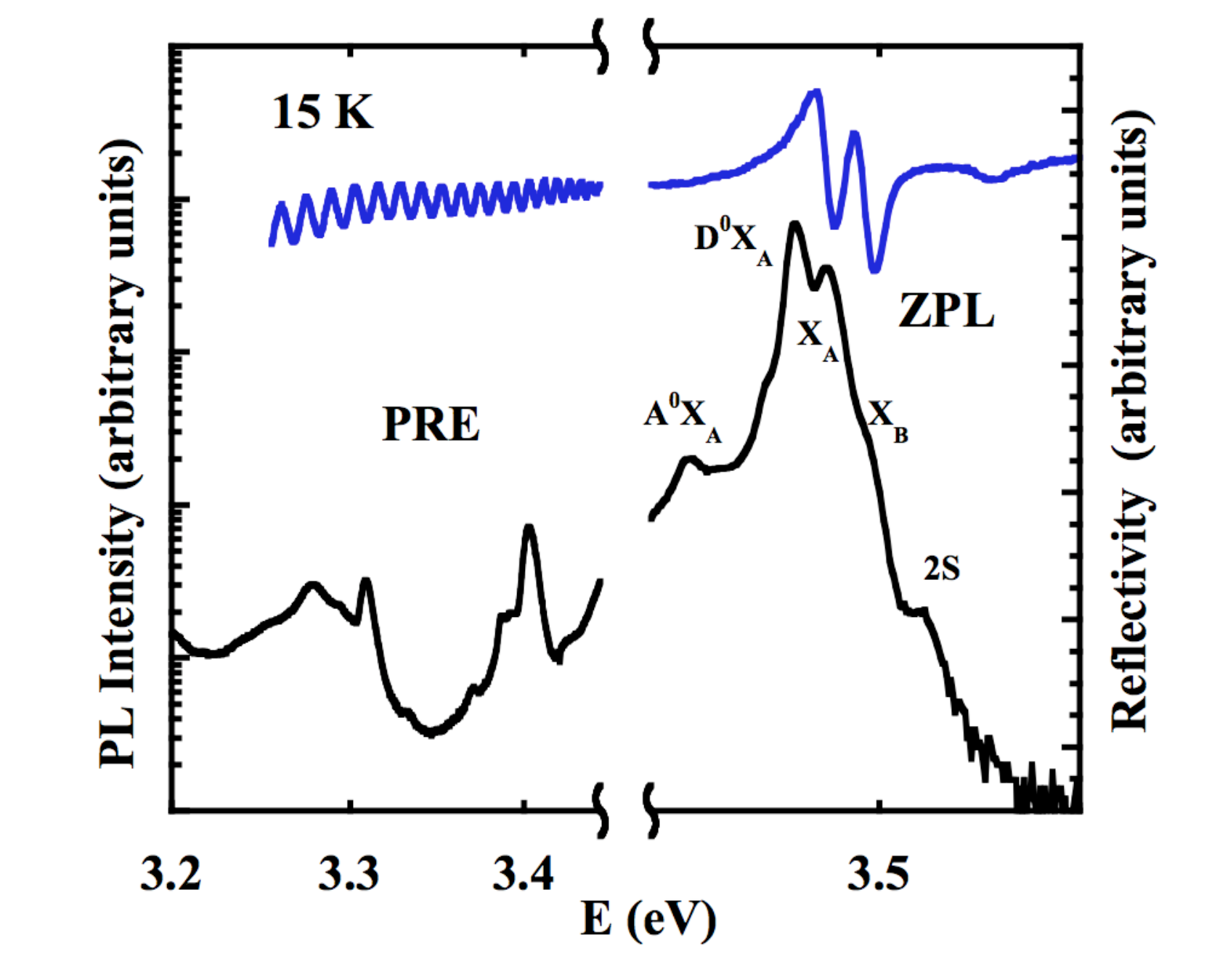}
	\caption{PL and reflectivity spectrum measured at 15K showing the details of the ZPL and phonon  replica (PRE)  bands}
	\label{Fig1}
\end{figure}
 In Fig. \ref{Fig1} typical TI-PL spectra taken at $T=15$ K for the ZPL and the PR region are shown for sample A. Similar spectra for sample B are shown in Ref. \onlinecite{cavigli2010}.  The main spectral features observed in the ZPL region at 10 K are related to the neutral donor-bound exciton recombination (D$^0$X$_A$) and to the X$_A$ and X$_B$ exciton radiative recombinations. The emission from the 2S state of X$_A$ excitons is also detected at higher energy. The reflectivity spectrum (blue line in Fig. \ref{Fig1}) shows clear excitonic resonances confirming the labeling of the PL structures. In this case the X$_C$ excitonic resonance is detected at $\sim$ 3.52 eV. In the PR region the reflectivity spectrum shows only interference fringes  related to the epilayer thickness. From the comparison of Fig. \ref{Fig1} and the spectra reported in Ref. \onlinecite{cavigli2010} it turns out that samples A and B show very similar features and pronounced excitonic resonances. The measured Stokes shift for the X$_A$ exciton, detected at 15 K, is also very similar and is lower than 1.5 meV in both samples indicating a weak exciton localization at low $T$. The difference observed in the exciton energies, as extracted from the reflectivity spectra, comes from  the different residual biaxial strains of the two epilayers and agrees with data and calculation already published.\cite{Shan96,gil97}  For a detailed discussion on the TI-PL spectra we refer the readers to Ref. \onlinecite{cavigli2010}. We only mention that the TI-PL lineshape almost coincides with the long delay TR-PL lineshape and that it well agrees with the thermal model of Segall-Mahan\cite{Segall68} when the discrepancy commonly reported for the temperature behavior of the phonon-replicas is described by accounting for the complexity of the GaN excitonic band.\cite{cavigli2010}  Here we will focus on the time-resolved spectra to investigate the typical time scale for reaching a quasi-thermal distribution for which the model of Ref. \onlinecite{Segall68} can be applied. Since the experimental results turn out to be very similar for samples A and B , most of the data we will show will refer to sample A. 
\begin{figure*}
\includegraphics[scale=0.65]{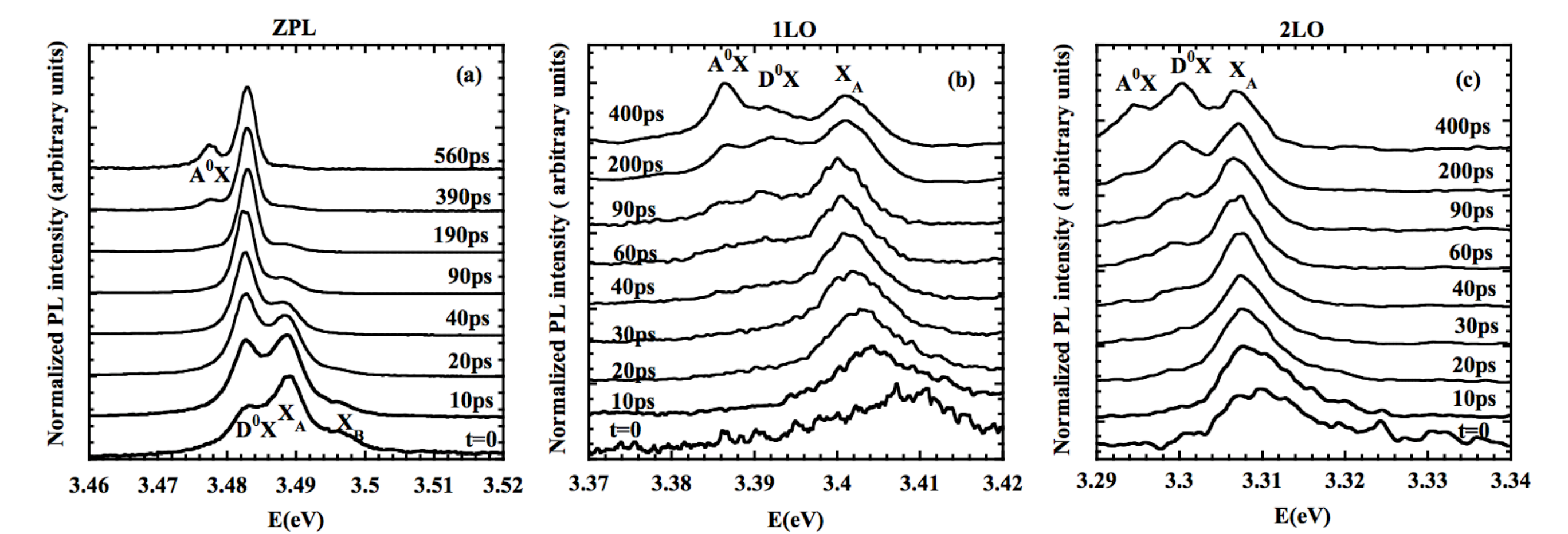}% Here is how to import EPS art
\caption{TR spectra of the (a) ZPL, (b) 1LO-phonon and (c) 2LO-phonon replica for different delay times after the excitation.}
\label{Fig2} 
\end{figure*}
In Fig. \ref{Fig2}, typical TR-spectra are shown for (a) the ZPL, (b) the first LO phonon replica (1LO) and (c) the second LO phonon replica (2LO) acquired when exciting the sample with 3.55 eV photons (i.e., photons with an excess energy of nearly 60 and 35 meV with respect to the bottom of the exciton and free carrier band, respectively, which is however still lower than the energy $E_{LO} = 92$ meV of the dominant LO phonon). Here, and in the following, the zero time corresponds to the maximum of the excitation pulse as measured from the Rayleigh or Raman scattered light. It appears that for the ZPL most of the dynamics involving free exciton states, both A and B ones, occurs during the first 40 ps following the excitation.  The modifications seen at later times in the ZPL lineshape come from the increasing contribution of the donor- and acceptor- bound exciton states. Concerning the time evolution of the replica bands we will first focus on the 2LO-phonon band whose shape and intensity is directly proportional to the exciton population along the k-states.\cite{Segall68}  At early times, the emission is blue-shifted and broadened with respect to that measured at delays longer than 100 ps.   Besides, for delays longer than 40 ps, the spectrum  gets narrower and shifts to the energy position expected for a thermal exciton population. 
\begin{figure}
\includegraphics[scale=0.6]{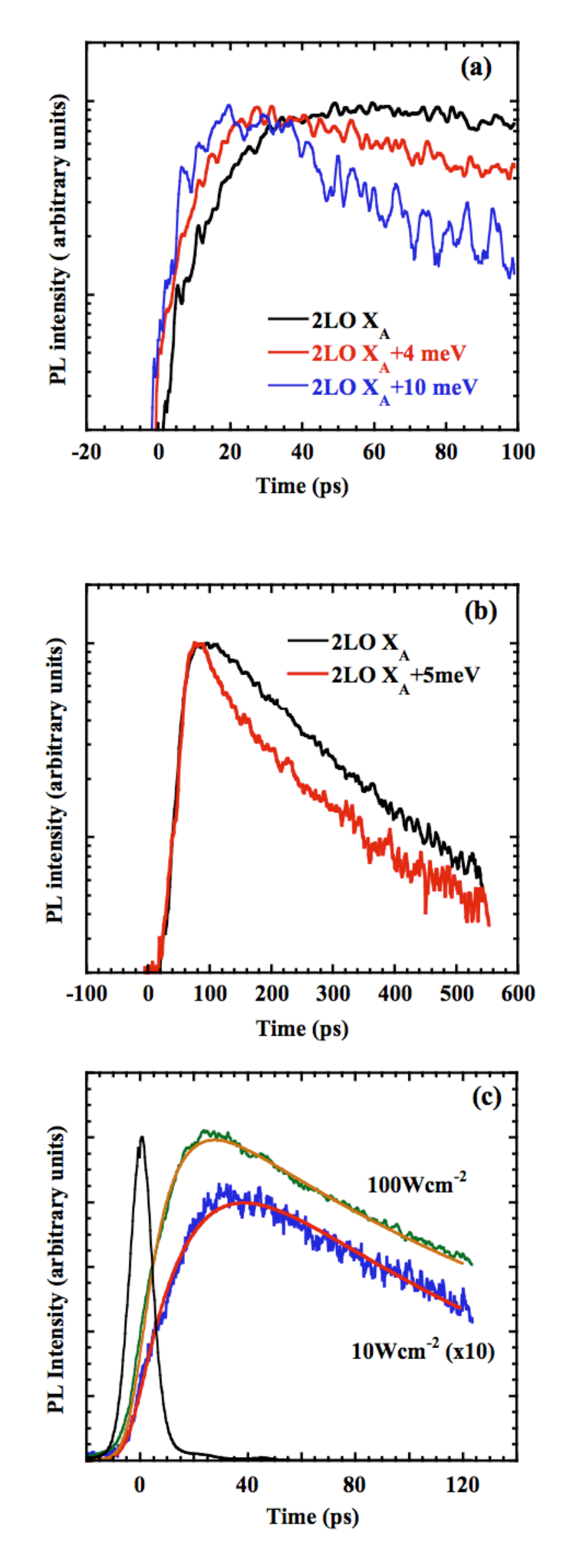}
\caption{PL time evolution at different energies around the 2LO-phonon  X$_A$ exciton (a) in a restricted and (b) over an extended time scale. (c) Time evolution of the spectrally integrated emission (only the free exciton contribution) of the 2LO-phonon band compared with the experimental time response (black line) for two different excitation densities. Orange and red lines are fits as discussed in the text.}
\label{Fig3}
\end{figure} 
The complexity of the exciton dynamics, which involves a non-thermal and a quasi-thermal regime at the picosecond time scale, is suggested by the different time evolutions of the spectral components of the 2LO-phonon band reported in Fig. \ref{Fig3}(a) and Fig. \ref{Fig3}(b), respectively. As a general trend, the rise and fall time get shorter with increasing energy of the recombining state. The risetime of the high energy states (where only FE are present), at the limit of our experimental time resolution, indicates an exciton-formation time $\approx$10 ps followed by a PL decay occurring on a 100 ps time scale, which originates from both the population thermalization and decay. At early times the high energy side of the phonon replica is faster than the X$_{\mathrm{A}}$ emission and we recover a common exponential decay at longer delays, indicating the establishment of the thermal  regime. The lower energy states exhibit a slower rise and decay as a consequence of the population dynamics and/or the  exciton cooling. 
\\In Fig. \ref{Fig3}(c), we show the time evolution of the spectrally integrated free exciton emission of the 2LO-phonon band, excited with an excess energy of $\approx$60 meV for two different power densities  (10 Wcm$^{-2}$ and 100 Wcm$^{-2}$) normalized to the excitation power density. The fit with an exponential rise and decay is reported on the experimental data and, for clarity, the experimental time response is also shown. In this case the rise and decay times reflect the time evolution of the total exciton population free from thermalization effects. In the following we will refer to the rise-time of the spectrally integrated PL as to the "exciton-formation time". Figure \ref{Fig3} (c) clearly shows a decrease in the exciton formation time (and a 20\% superlinearity) when increasing the excitation power density from 10 to 100 Wcm$^{-2}$. In addition the exciton formation time only sligthly depends on the excitation energy as supported by the decrease from 13 to 9 ps when reducing the excitation excess energy below the free carrier absorption edge (from 60 to 20 meV with respect to the bottom of the exciton band). This suggests, as we will discuss later, that we mostly create electron-hole (e-h) correlated pairs rather than free carriers, even when we excite the samples a few tens of meV above the free carrier continuum. 
\begin{figure}
 \includegraphics[scale=0.35]{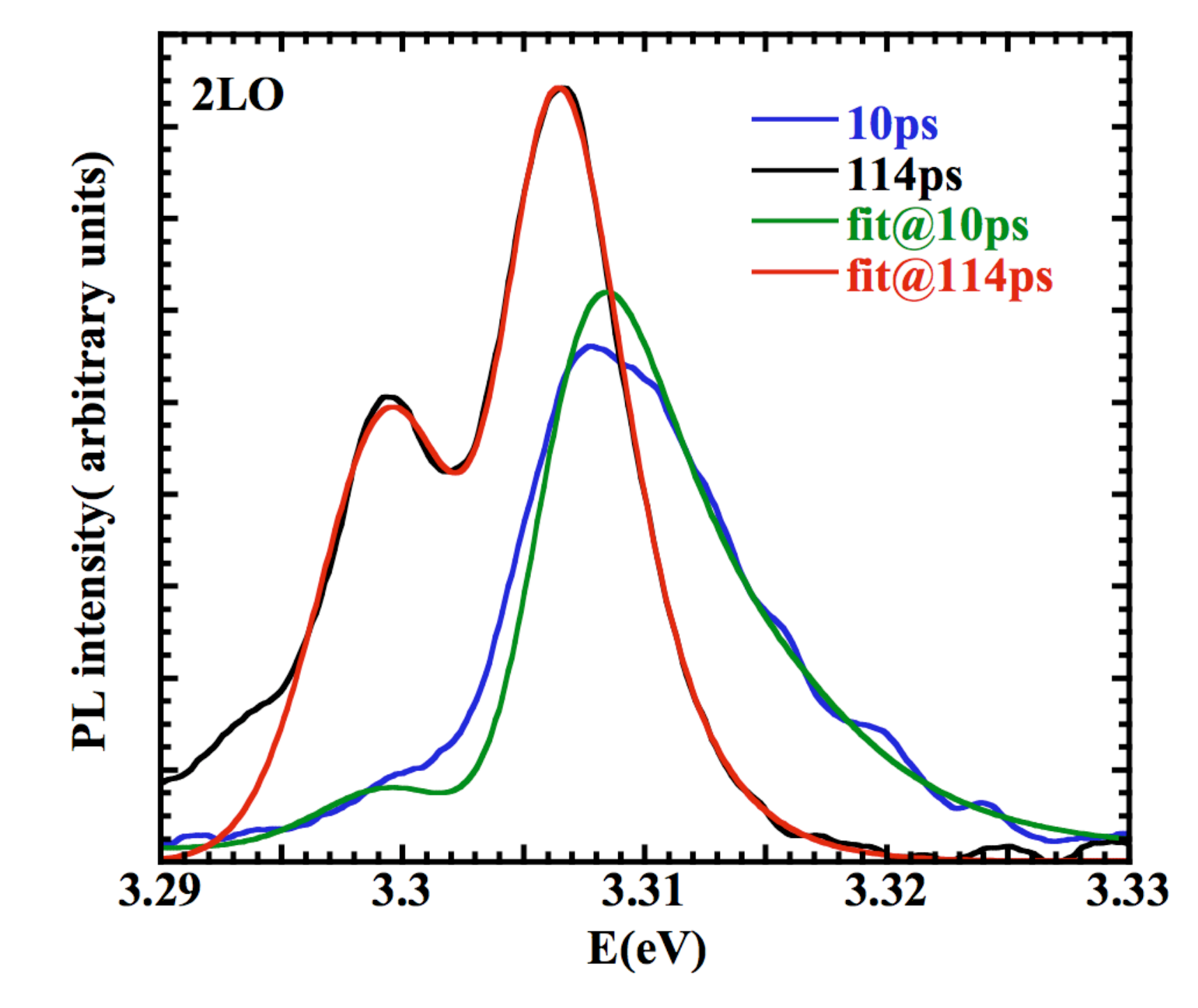}
 \caption{TR spectra of the 2LO-phonon band at different delays along with a fit as discussed in the text.}
 \label{Fig4}
 \end{figure} 
\\Let us now move to the study of the exciton relaxation process. In Fig. \ref{Fig4} we show the comparison between the 2LO phonon replica time-resolved spectra taken at $t = 114$ ps and $t = 10$ ps delay time after the excitation pulse. It is worth pointing out that the overall exciton dynamics is almost independent from the excess energy of the exciting photons (not shown here); therefore in the following the experimental data will refer to the excitation energy of 3.55 eV. The exciton relaxation dynamics has been commonly interpreted as a cooling process assuming that an exciton thermal distribution can be reached within a few picoseconds\cite{Hagele99,Shah96} and then that we can reproduce the PL lineshape with a Boltzmann distribution with a temperature dependent on the delay time. In fact, in agreement with previous reports,\cite{Hagele99} we find that, within the experimental sensitivity, the high energy tail of the emission is mostly exponential even at very short delay times. However, we can perform a more accurate analysis by using the fact that the lineshape of the 2LO phonon band is directly related to the population along the exciton band by
\begin{equation}
I_{2\mathrm{LO}}(E,t)\varpropto\rho(\epsilon)f(\epsilon,t)\Theta(E-E_0)
\label{eq1}
\end{equation}
where $\epsilon=(E-E_0)$ is the exciton kinetic energy, $E_0=E_{X_A}-2E_{LO}$, $\rho(\epsilon)$ represents the exciton density of states and $f(\epsilon ,t)$ is the exciton distribution function at time $t$. Generally speaking, $f(\epsilon,t)$ can even be a non-thermal distribution and, in fact, reasonable fits to the time-resolved 2LO-phonon emission spectra with Boltzmann functions for $f(\epsilon,t)$ are obtained only at long delays. In Fig. \ref{Fig4}, we compare the time-resolved 2LO-phonon  spectra with the corresponding fits using a Boltzmann-like distribution. The spectrum at a delay time of 114 ps is nicely reproduced by $\rho(\epsilon)$ given in Ref. \onlinecite{cavigli2010} and $k_BT_L$ = 1.97 meV, $T_L$ being the temperature of the lattice. The best fit to the data at a delay time of 10 ps requires a hotter thermal distribution ($k_BT$ = 3.77 meV). However the quality of the fit is quite poor. In particular the states near the peak energy  reveal a lack of population with respect to the thermal condition, suggesting a non-thermal exciton distribution. 
In order to quantitatively discuss the exciton thermalization we, first of all, normalize each time-resolved spectrum of the 2LO-phonon PL band with its spectrally integrated intensity. This operation eliminates  the exciton formation and recombination from the lineshape analysis, which, to a first approximation, only affects the total number of excitons. The resulting normalized spectra $I^*_{2\mathrm{LO}}(E,t)$ is then proportional to the normalized exciton population $N(E,t)$. The next step is to bypass the problem of modelling $\rho(\epsilon)$. To this aim, considering the ratio $n(E,t)=I^*_{2\mathrm{LO}}(E,t)/I^*_{2\mathrm{LO}}(E,t=200 \mathrm{ps})$ between the normalized spectra, we have experimental data that do not depend on the actual $\rho(\epsilon)$. In fact,  according to the experimental data, we can assume a thermal distribution at the lattice temperature  for $t > 100$ ps ( hereafter $t =\infty$ ). Then from Eq.\ref{eq1} the 2LO-phonon normalized emission band at $t =\infty$   is given by:
\begin{equation}
I^*_{2\mathrm{LO}}(E,\infty)\varpropto\dfrac{\rho(\epsilon)e^\frac{-\epsilon}{k_BT_L}}{(k_BT_L)^{(3/2)}}\Theta(E-E_0)
\label{eq2}
\end{equation}
and therefore from $n(E,t)$ we can extract the exciton distribution $f(\epsilon ,t)$ even if the density of states is unknown.
In Figs. \ref{Fig5}(a) and \ref{Fig5}(b) we report the time evolution of the normalized spectra $I^*_{2\mathrm{LO}}(E,t)\propto N(E,t)$  at fixed energies for two different excitation power densities (10, 100 Wcm$^{-2}$). The quantity $N(E,t)$ decreases with time for large $E$ values, while it increases for small $E$ values. These trends are obviously associated with the transfer of population towards the bottom of the excitonic band. We also observe that the state of energy around 3.310 eV, nearly corresponding to an excess of energy $3k_BT_L/2$ above the bottom of the exciton band, rapidly reaches the equilibrium value at temperature $T_L$, i.e., a constant value of the population. A faster thermalization is observed at higher excitation power density, as expected if we take into account exciton-exciton interaction. 
In order to get more direct information on the exciton distribution $f(\epsilon,t)$ we define the quantity  $\beta^*(E,t)$:
\begin{eqnarray}
\beta^*(E,t)=\mathrm{ln}\frac{I^*_{2\mathrm{LO}}(E,t)}{I^*_{2\mathrm{LO}}(E,\infty)}=\nonumber \\
=\mathrm{ln}\left[(k_BT_L)^{(3/2)}f(\epsilon,t)\right]+\frac{\epsilon}{k_BT_L}
\label{eq3}
\end{eqnarray}
$\beta^*(E,t)$ turns out to be an indicator of the degree of thermalization of the excitons and it is independent on the actual density of states. Whenever the hot exciton picture applies, i.e., when the exciton distribution $f(\epsilon ,t)$ is a quasi-thermal distribution with a time-dependent temperature $T(t)$, we get 
\begin{eqnarray}
\beta^*(E,t)_{th}=\epsilon\left(\frac{1}{k_BT_L}-\frac{1}{k_BT(t)}\right)+\frac{3}{2}\mathrm{ln}\left( \frac{k_BT_L}{k_BT(t)}\right) 
\label{eq4}
\end{eqnarray}
In the hot exciton limit we therefore expect to get a straight line when plotting  $\beta^*(E,t)$ as a function of energy with a slope given by $1/k_BT(t)$. Any deviation from a linear dependence vs. $E$ is therefore a signature for a non-thermal regime. In Fig. \ref{Fig6} (a),  $\beta^*(E,t)$ is reported for different delay times, in the 2LO-phonon spectral region, as extracted from TR spectra. The nominal lattice temperature is equal to 10 K and the sample is excited with an excess energy of $\approx60$ meV with respect to $E_0$ with  a power density of 10 W cm$^{-2}$. For clarity we report the value of $\beta^*(E,t)$ only in the spectral region of the free exciton emission with the low energy side cutoff  corresponding to the onset of the 2LO-phonon D$^0$X$_{\mathrm{A}}$ emission. In this figure, the energy $E$ is identified with the photon emission energy and not only the kinetic energy term.  We also show the time-integrated PL spectrum of the 2LO-phonon replica corresponding to the full thermal regime. Strong deviations from a linear trend of  $\beta^*(E,t)$ are observed at the early stage of the exciton dynamics ($t < 40$ ps). In particular a decrease in the exciton population for small kinetic energies with respect to a thermal distribution is found. In addition, we also observe a flattening of $\beta^*(E,t)$ on the high energy side (corresponding to the tail of the PL band), which could be a further indication for a non-thermal distribution. However, this part of the spectra is affected by a relevant background whose subtraction is delicate making reliable conclusions rather hazardous. It turns out that the distribution function behaves like a Boltzmann function only at longer delays ($t > 50$ ps) and  $\beta^*(E,t)$ shows a linear dependence over the whole energy range covered by the 2LO-phonon band with a decreasing slope in agreement with a cooling of the exciton population. Finally  $\beta^*(E,t)$ becomes constant and vanishes, as expected, when the distribution reaches the lattice temperature. Referring to Fig. \ref{Fig6}(a), the deviation of $\beta^*(E,t)$ from the linear behavior observed for $t < 40$ ps indicates the lack of excitonic population not only with respect to the thermal regime at $T = T_L$ but also with respect to a hot exciton distribution. Therefore our findings clearly evidence the presence of a non-thermal distribution of the exciton population at short delay times. It comes out that the cooling picture of a hot exciton distribution does not apply if we refer to the whole excitonic band. Moreover, the lack of population at the X$_{\mathrm{A}}$ energy with respect to the hot thermal distribution cannot be ascribed to a localization mechanism. In fact, the absence of any significant localization of the excitons is proven by the comparison of reflectivity and PL spectra for the ZPL (Fig. \ref{Fig1}) where it turns out that the Stokes shift at 10 K is $\approx 1$ meV, a value not large enough to explain the experimental results displayed in Fig. \ref{Fig6}(a). In Fig. \ref{Fig6}(b), similar spectra are reported for an excitation power density of  100 Wcm$^{-2}$. No major change is found except that the quasi-thermal regime is reached for slightly shorter times with respect to the low excitation condition, as already shown in Fig. \ref{Fig5}. Up to 50 K the non-thermal regime can still be detected during the first 30 ps and similar results for $\beta^*(E,t)$ are observed when we change the photon excitation energy  for an excess energy lying in the 6 to 60 meV range. 
\begin{figure}
 \includegraphics[scale=0.32]{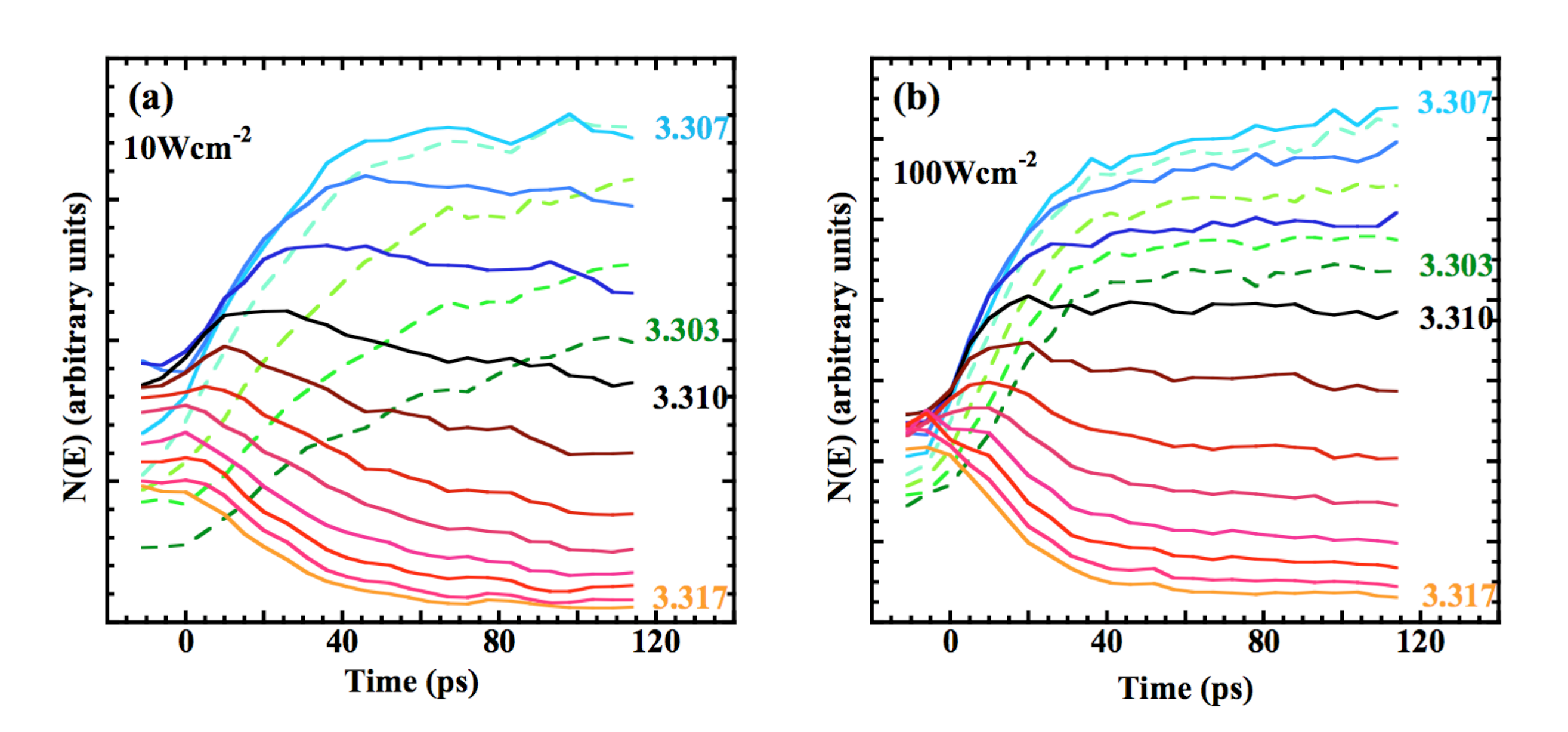}
 \caption{Time evolution of the normalized exciton population N(E) at different energies (equi-spaced) in the 2LO band for  10 Wcm$^{-2}$ (a) and 100 Wcm$^{-2}$ (b) excitation power densities. The continuous (dashed) lines refer to states lying above (below) 3.307 eV.}
 \label{Fig5}
 \end{figure}
A specular behavior is instead observed when we resonantly excite the k=0 X$_{\mathrm{A}}$ exciton and we study the heating of the corresponding exciton distribution. In Fig. \ref{Fig7}(a) TR spectra are shown for different delay times after the resonant excitation at E = 3.303 eV slightly below the X$_{\mathrm{A}}$ exciton.  Apart from the initial Raman signal detected for the first 10 ps, that dominates the emission spectra at the early stage, the exciton line and also that of the D$^0$X$_{\mathrm{A}}$ emission are visible at $t\geq20$ ps. In Fig. \ref{Fig7}(b), we report $\beta^*(E,t)$ for $t \geq 20$ ps when the Raman signal can be neglected and we can discuss the evolution of the exciton distribution without spurious emission. Due to this cut occurring at the early stage in the time evolution of the excitons it is not possible to have a definitive proof of their non-thermal distribution under this excitation condition. Still $\beta^*(E,t)$ shows a slope opposite to that reported for non-resonant excitation, denoting a cold exciton distribution. During the thermalization process we observe an exciton heating with a population transfer between low and high energy states until the equilibrium distribution at temperature $T_L$ is reached. Quite remarkably we find that the time scale for reaching the equilibrium distribution is almost the same for the cooling and heating of  the exciton population. Finally it is worth noting that, when comparing the PL spectra shown in  Figs. \ref{Fig6} and \ref{Fig7} a saturation of the acceptor emission is observed (Fig. \ref{Fig6}) when increasing the excitation power density while it is its absence which is observed when exciting the sample resonantly (Fig. \ref{Fig7}).
\begin{figure}
 \includegraphics[scale=0.45]{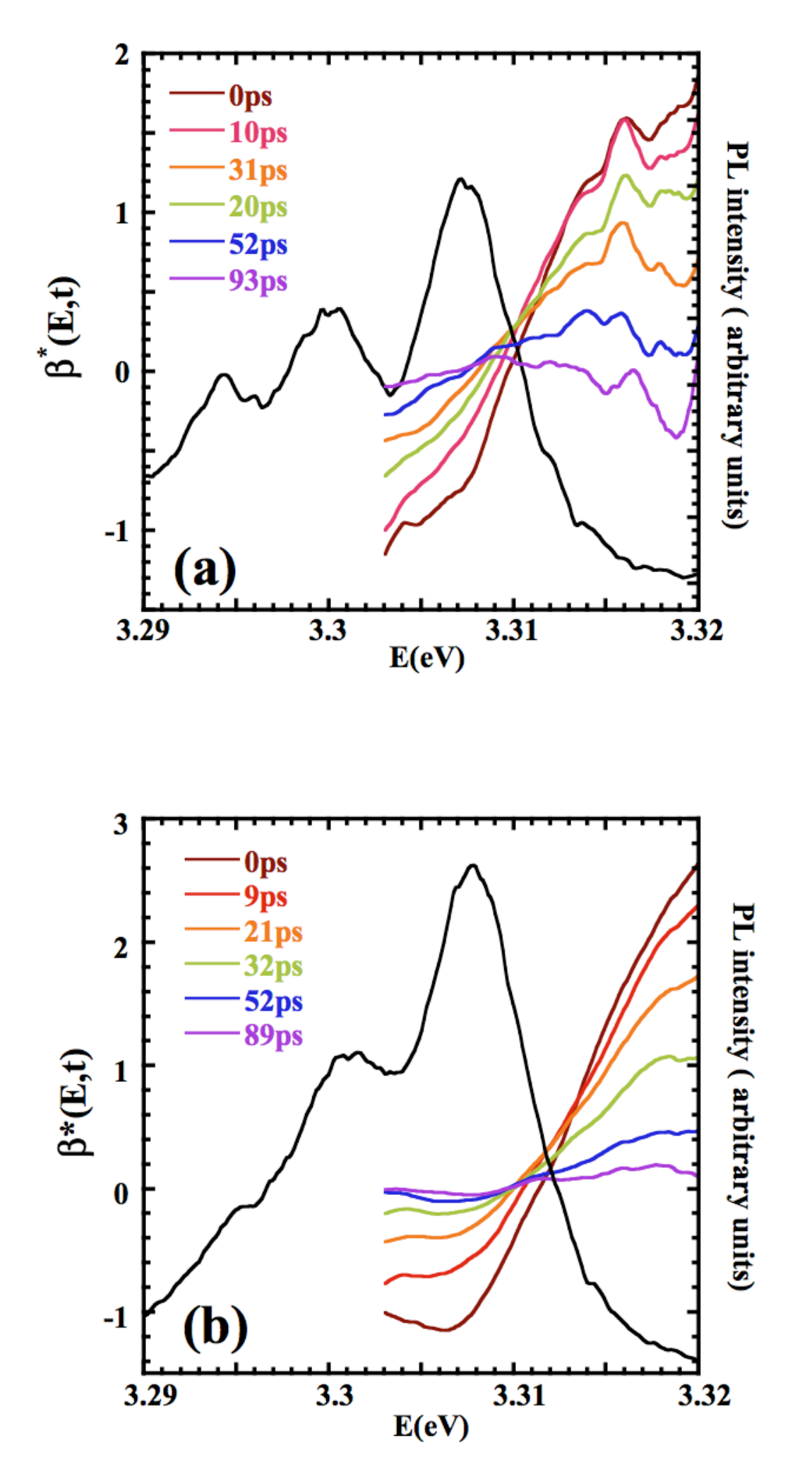}
 \caption{Time evolution of the quantity $\beta^*(\epsilon,t)$ for different delay times for non-resonant excitation:  10 Wcm$^{-2}$ (a) and 100 Wcm$^{-2}$ (b). The solid black line in (a) and (b) is the time-integrated PL spectrum.}
 \label{Fig6}
 \end{figure}
 
\begin{figure}
 \includegraphics[scale=0.4]{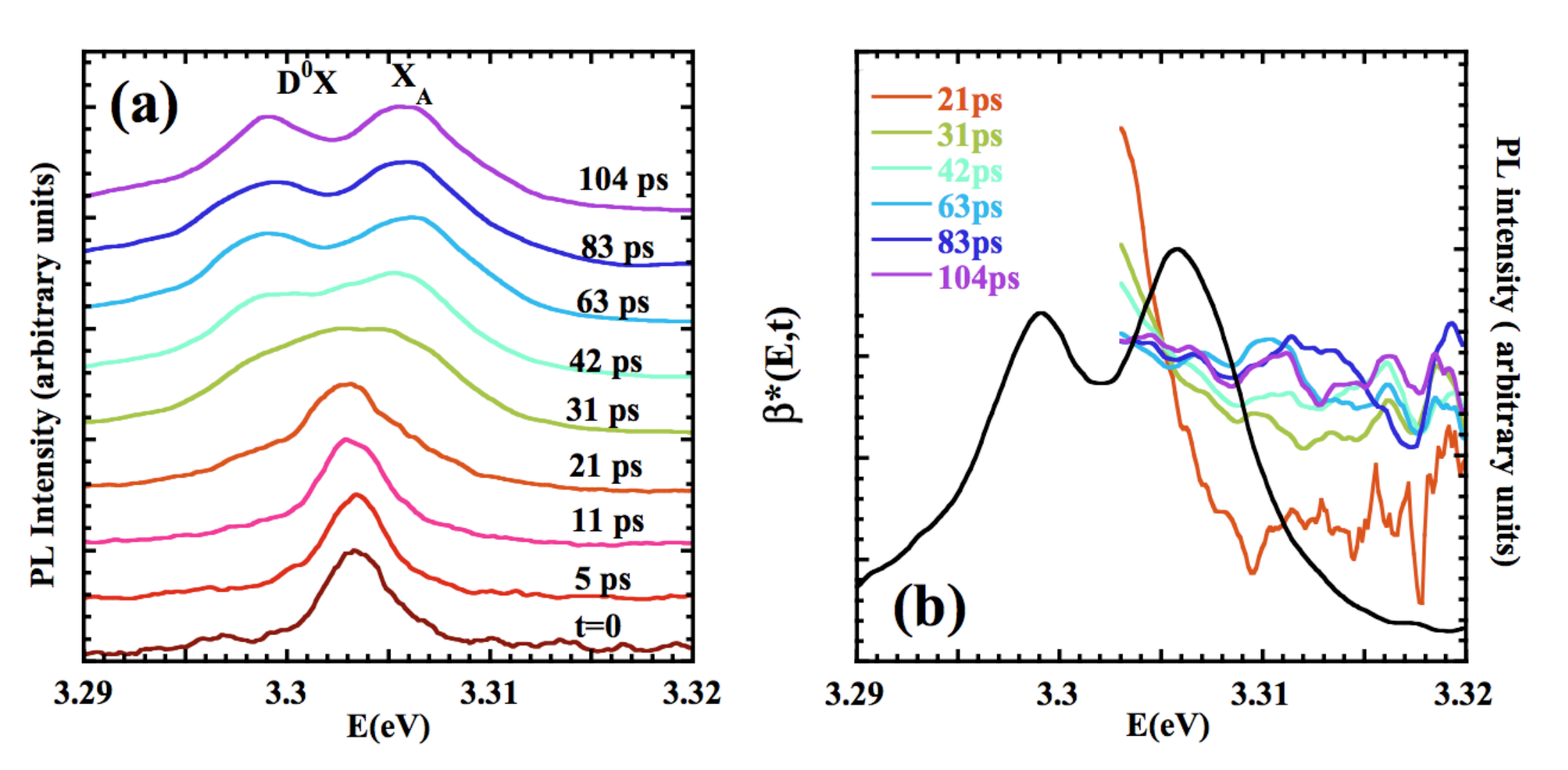}
 \caption{(a) Time resolved spectra taken at different delay times for resonant excitation. (b) Time evolution of the quantity $\beta^*(\epsilon,t)$ for the resonant case. For clarity the 2LO-phonon PL spectrum in the fully thermalized regime is also shown (black solid line).}
 \label{Fig7}
 \end{figure}
\section{Discussion}
\begin{figure}
\includegraphics[scale=0.6]{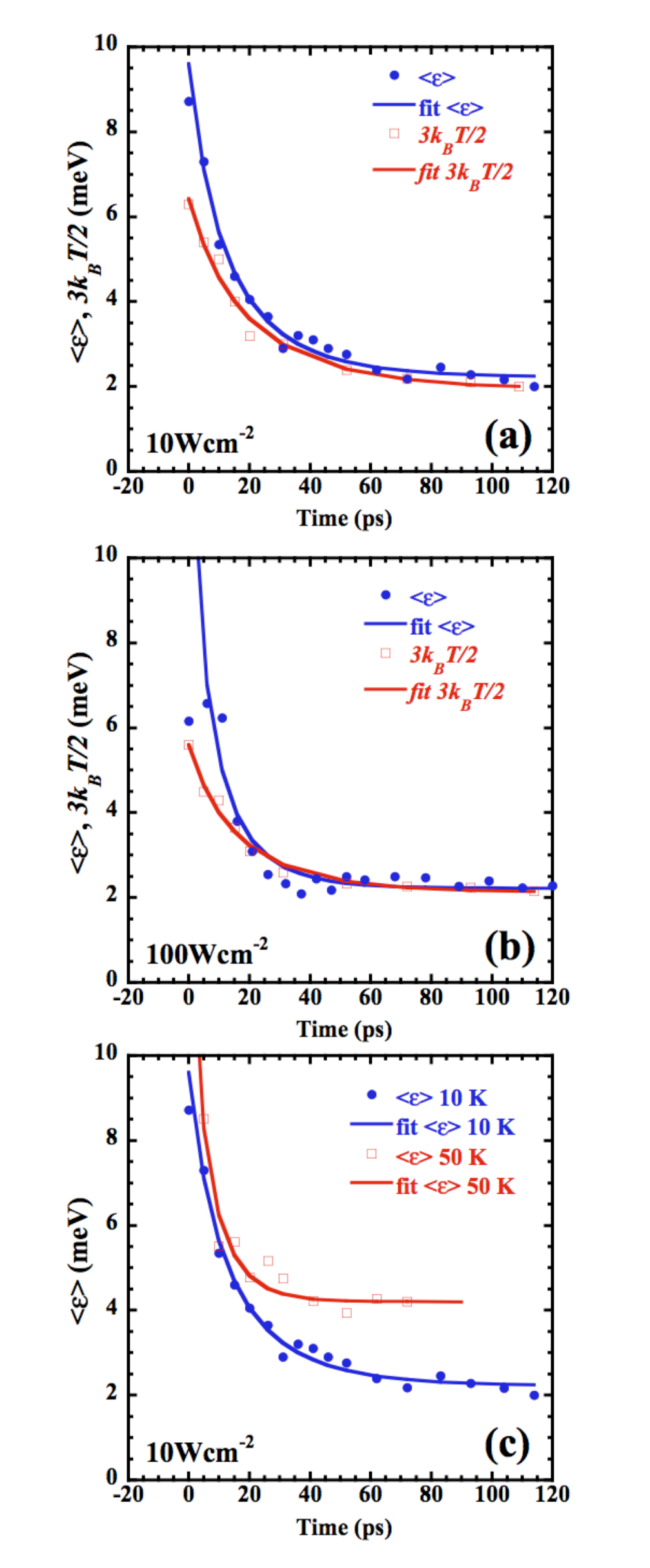}
 \caption{Time evolution of the exciton average kinetic energy $<\epsilon >$ and the thermal energy along with the fits as discussed in the text for the non-resonant excitation case. (a) 10 Wcm$^{-2}$ . (b) 100 Wcm$^{-2}$ .(c) 10 K and 50 K, nominal lattice temperature at 10 Wcm$^{-2}$ .}
 \label{Fig8}
 \end{figure}
The analysis of the 2LO PR-PL lineshape, without any supplementary modeling of the exciton density of states, already gives important information on the excitonic distribution function. However, to quantitatively study the thermalization process, we have to analyze and compare the time evolution of the mean energy $<\epsilon (t)>$ and that of the effective temperature $T(t)$ of the exciton distribution.  $<\epsilon (t)>$ can be directly obtained from the experimental data by considering the ratio between the spectrally integrated 1LO-phonon emission and the 2LO-phonon band,\cite{ulbrich73} while $T(t)$, according to the usual approach,\cite{Hagele99} is given by an exponential fit of the high energy tail of the 2LO-phonon PL spectra. Obviously for a thermal exciton distribution and a standard excitonic density of states we expect to have $<\epsilon (t)> =3k_BT(t)/2$. The two data sets, $<\epsilon (t)>$ and $3k_BT(t)/2$, are reported in Figs. \ref{Fig8}(a) and \ref{Fig8}(b) for two excitation power densities. We observe that in the early stage $<\epsilon (t)>$ is larger than $3k_BT(t)/2$, thus denoting a non-thermal distribution of the excitons. At long delay time $<\epsilon (t)>$ and $3k_BT(t)/2$ reach the same asymptotic values, as expected for an excitonic thermal distribution. In Fig. \ref{Fig8}(c) the value of $<\epsilon (t)>$ is reported for two different lattice temperatures $T_L= 10$ and 50 K. At $T_L = 50$ K we see that the exciton cooling is significantly faster than that at $T_L = 10$ K. In Figs. \ref{Fig8}(a)-\ref{Fig8}(c) the continuous lines are best fits to the experimental data obtained with the expression for the time-dependent mean energy $<\epsilon (t)>$ (or $3k_BT(t)/2$):
\begin{eqnarray}
<\epsilon(t)>=<\epsilon_{\infty}>\left(\dfrac{Ae^{t/\tau}+1}{Ae^{t/\tau}-1} \right)^2 
\label{eq5}
\end{eqnarray}
with
\begin{eqnarray}
A=\frac{\sqrt{<\epsilon_0>}+\sqrt{<\epsilon_{\infty}>}}{\sqrt{<\epsilon_0>}-\sqrt{<\epsilon_{\infty}>}}
\label{eq6}
\end{eqnarray}
where  $<\epsilon_{\infty}>$ and $<\epsilon_0>$  are  the equilibrium and initial (i.e. at $t = 0$) mean energy .  The thermalization time constant $\tau$ is determined by the exciton-acoustic phonon interaction due to the deformation potential\cite{ulbrich73,Hagele99} and does not depend on the lattice temperature $T_L$. This expression has been very often derived assuming a thermalized distribution, therefore referring to the temperature. Nevertheless it holds independently from this assumption.\cite{Ridley88} The values determined for the time constant of  $<\epsilon>$  and  $3k_BT(t)/2$  are respectively $\tau = 23\pm2$ ps and  $32\pm2$ ps  at 10 Wcm$^{-2}$ and  $\tau = 15\pm1$ ps and  $24\pm1$ ps at 100 Wcm$^{-2}$; for $T_L = 50$ K  we get  $\tau = 12\pm1$  ps.
As expected the cooling process becomes faster when rising either the excitation power density or the temperature. In addition the data fitting points out a strong discrepancy between $<\epsilon>$ and $3k_BT(t)/2$, strenghtening the non-thermal distribution character of the excitonic gas. 
\\We can now compare the energy relaxation time with the values predicted by theories. From  $\tau =  23$ ps  (32 ps), by means of the expression reported in Ref. \onlinecite{ulbrich73}, we  get an effective deformation potential  (DP) constant  E$_1 = 20 \pm1$ eV  ($17 \pm1$ eV) to be compared with the value E$_1 =  13 \pm1$ eV reported in Ref. \onlinecite{Hagele99} and E$_1 =  9 \pm2 $eV  usually found both theoretically and experimentally (e.g. see Ref. \onlinecite{Ishii2010}).  Provided the value we find is nearly twice the "standard" one, and that the latter would lead to a thermalization time of about 100 ps, we should consider other scattering mechanisms in addition to that of the deformation potential. A contribution to the exciton thermalization from exciton-exciton scattering is evidentiated by the acceleration of the dynamics when rising the excitation power density from 10 to 100 Wcm$^{-2}$, i.e the exciton density from $\approx 10^{16}$ cm$^{-3}$ to $\approx 10^{17}$ cm$^{-3}$. In fact the role of the inelastic exciton-exciton collision in the population redistribution in GaN has  been reported by several autors.\cite{bidnyk99,tanaka06} An evaluation of the exciton-exciton scattering rate can be obtained by the dephasing time. Indeed, according to the results of Pau \textit{et al.} (see Ref. \onlinecite{Pau97}) this contribution to the linewidth depends on the exciton density N$_{\mathrm{X}}$ as $\Gamma_{\mathrm{XX}} =  \beta _{\mathrm{XX}}\mathrm{a}_B^3 E_B \mathrm{N}_{\mathrm{X}}$ , where  a$_B$ is the exciton Bohr radius, $E_B$ the exciton binding energy and $\beta_{\mathrm{XX}}$ a dimensionless efficiency factor which for GaN is between 5 and 7.  With a$_B =  3$ nm  and   $E_B = 24$ meV, as found from an hydrogenoid model,  we get a dephasing time $T_2 = \pi \Gamma_{\mathrm{XX}}/h$  in the range  20 to 40 ps   for  N$_{\mathrm{X}} =  10^{16}$ cm$^{-3}$. We get instead a value of 2 to 4 ps   for  N$_{\mathrm{X}} =  10^{17}$ cm$^{-3}$, which is an order of magnitude too short to account for the experimental results. From our data, it turns out that, in the thermalization process, the inelastic exciton-exciton scattering contributes but with an efficiency ten times lower than in the dephasing process, where collisions are prevailing without any change in the energy of the single exciton. It is also worth pointing out that the exciton-free carrier scattering  process can in principle also contribute to the exciton relaxation but, from time-resolved spectra, the number of free carriers seems negligible with respect to that of the excitons.   Therefore other mechanisms have to be considered and in particular the piezoelectric interaction (PZ) with the acoustic phonons as well as the impurity (defects) inelastic (D$^0$X$_{\mathrm{A}}$) scattering. The role of the PZ interaction in the exciton thermalization process in GaN epilayers has been extensively investigated by Kokolakis \textit{et al}. (see Ref. \onlinecite{Kokolakis03})and it has been found nearly equal to that of the DP interaction for low residual doping. Due to screening effects it becomes negligible for a $10^{17}$ cm$^{-3}$  impurity content.  Provided that the PZ mechanism does not give a good fit to the experimental data we consider its contribution to the thermalization as being marginal. When considering the scattering from neutral donors the impact rate is roughly given by $\Gamma_{\mathrm{DX}} = \mathrm{a}_B^2\mathrm{v_gN_D}$,\cite{wang09,Steiner86} where N$_\mathrm{D}$ is the neutral donor concentration and v$_\mathrm{g} = \surd2\epsilon/\mathrm{m}$ is exciton group velocity, being $\epsilon$ and m the exciton  kinetic energy and mass. Then, for example, with $\epsilon = 2$ meV  and N$_\mathrm{D} =  10^{17}$ cm$^{-3}$  we get a very short collision time $\tau_\mathrm{DX} = (\Gamma_{\mathrm{DX}})^{-1} \approx400$ fs. Therefore, even if most of the collisions are of elastic nature and that they only contribute to the exciton dephasing process,\cite{wang09}  we can expect a relevant contribution to the exciton thermalization from the inelastic processes as already found in GaAs\cite{Steiner86} and InP.\cite{Benzaquen73} This mechanism is also in agreement with the observed behavior when changing the lattice temperature $T_L$ provided the scattering rate only depends on the exciton energy and not on $T_L$. Thus, as observed in Fig. \ref{Fig8}(c), the initial energy relaxation towards the exciton thermalization is independent of $T_L$. Since the exciton initial excess energy is identical for $T = 10$ and 50 K, the acceleration observed in Fig. \ref{Fig8}(c) is only becoming apparent as a consequence of the higher final energy of the excitons at 50 K. This does not hold for the DP (PZ) interaction. In fact the fit of the experimental data of Fig. \ref{Fig8}(c) with Eq. \ref{eq5} requires a decrease in the time constant   related to the DP (PZ) interaction, when rising the temperature, in disagreement  with its expression, which is independent of $T_L$.\cite{ulbrich73,Hagele99}
\\The previous discussion deals with the time evolution of a hot exciton population that has reached a finite and constant value and evolves towards the thermal equilibrium with the lattice. Let us now briefly comment on the early stage of the exciton dynamics when the states near the bottom of the excitonic band become populated, i.e., the exciton formation stage, which in our measurements corresponds to the time interval between the arrival of the excitation pulse and the observation of the maximum of the spectrally integrated PL of the 2LO-phonon replica(Fig. \ref{Fig3} (c)).  The  corresponding PL rise time turns out to be $\approx 10$ ps,  nearly independent on the energy of the excitation photon, which is only slightly shortened when rising the excitation power density. When exciting the sample $\approx 60$ meV above the  X$_A$ state, i.e 35 meV above the free carrier continuum, we create both free carriers and correlated e-h pairs. Free carriers relax in energy and form excitons via a bimolecular (BM) mechanism with a coefficient  $\mathrm{B} = 1.2\times 10^{-18}$ ps$^{-1}$ cm$^3$.\cite{PhysRevB.58.12916} Correlated e-h pairs reach the bottom of the exciton band losing their excess energy by means of inelastic interactions with acoustic phonons, impurities and defects. At the excitation power densities used in our experiment the bimolecular generation only contributes at the highest power density (100 Wcm$^{-2}$), i.e. $10^{17}$  carrier/cm$^3$),providing an exciton formation time of 12 ps, not far away from the observed rise time of the 2LO-phonon replica. In addition the slight superlinearity shown in Fig. \ref{Fig3}(c) well agrees with the BM mechanism. In our experiment at  10 Wcm$^{-2}$ the exciton formation time is only sligthly longer and is still in the range of 10 ps. Then it turns out to be shorter than the time needed to reach the bottom of the exciton band when an energy loss of 30-40 meV by emitting acoustic phonons is considered. Moreover the BM generation would provide a risetime of $\approx 100$ ps.\cite{PSSA:PSSA141}  Therefore, also for the exciton formation, inelastic collisions with neutral donors appear as the most likely process for a fast exciton formation, or more precisely, for a fast relaxation of correlated e-h pairs towards the bottom of the exciton band. One inelastic collision, giving rise to the donor  ionization, is sufficient to produce a 20-30 meV loss in the exciton energy, thus providing a fast transfer from the high to the low energy exciton states. The fast rise in the ZPL signal, equal to that reported for the 2LO-phonon PL in Fig. \ref{Fig2} at low excitation power density is mainly due to exciton-impurity scattering, which avoids the bottleneck resulting from the reduced efficiency for the exciton-acoustic phonon scattering. 

\section{Conclusions}
From a careful analysis of low temperature time-resolved spectra of the phonon replica emission in GaN epilayers we have evidenced the transition from the non-thermal to the thermal regime of the exciton population. Our experimental results provide clear signatures that the processes leading to the exciton formation and thermalization are relatively slow when the exciton density is lower than $10^{17}$  cm$^{-3}$.  The establishment of a thermal Maxwell-Boltzman distribution for the exciton population is observed after $\approx 50$ ps,mainly as a consequence of inelastic collisions with impurities rather than with acoustic phonons.  From the comparison with models and predictions for the deformation potential, piezoelectric interaction and impurity scattering we show that neutral donor scattering can account for the exciton formation and relaxation. Only thanks to the residual doping,  the bottleneck regime for the radiative excitons at k = 0 is avoided. Nevertheless, for an exciton density $\geq 10^{17}$  cm$^{-3}$ , the exciton-exciton collisions become relevant in the exciton formation and thermalization, contributing to the cooling via  inelastic processes. 

\begin{acknowledgments}
The research was partially supported by the Italian Ministry of University and Research PRIN \textit{GaN optoelectronic devices for future applications: solid-state lighting, biomedical instrumentation and data storage}, 20095NMPW7.

\end{acknowledgments}

%\bibliography{biblionew.bib}% Produces the bibliography via BibTeX.
\providecommand{\noopsort}[1]{}\providecommand{\singleletter}[1]{#1}%

\end{document}